# (Re)conceptualizations:
# Intentional concept development in the social sciences

Piotr Tomasz Makowski

*Queen's Univesity Belfast & University of Warsaw*

p.makowski@qub.ac.uk

**Abstract**. Can intentional concept development in the social sciences be understood in terms of conceptual engineering (CE)? To answer this question, I analyze various types of conceptual changes in the social changes—with a special attention to organizational research and the so-called *(re)conceptualizations*—and distinguish between CE as a theoretical practice and CE as a research program. I show that social scientists, from the point of view of their scientific practice, exercise CE in two versions: CE de novo is employed as new *conceptualizations* and moderately progressive CE—as *reconceptualizations*. Importantly, the second type of CE—rather neglected in philosophy of the social sciences—appears to be highly important for the incremental progress of inquiry. Still, both types appear to be equally significant also for CE understood as a research program and for its prospects in the social sciences. Here, I point to three possible paths that help bridging the gap between actual practices of concept development in the social sciences and normative, programmatic approaches to CE: best practice recommendations, institutional actions and uses of AI-agents.

# *(Re)conceptualizations*:
# Intentional concept development in the social sciences

Planned manipulations and changes in the domain of concepts are not reserved for philosophers. Regardless their methods, scholars and scientists used to work a lot with concepts—or "constructs"—and much of what they do implies concept development. While such development isn't always palpable in hard sciences, it appears to be especially visible in the social sciences. In the latter, the dynamics of concepts (and inquiry) is significant due to the tempo with which human understanding changes together with the changeable social reality. This is why so called "theoretical contributions" often involve working with concepts: many papers turn out to be about concepts and their meaning. Changeability of social practices significantly contributes to the dynamics of inquiry which is one of the reasons why social and organizational concepts often require explicit attention in the soft sciences.

Concepts are formed anew with the aid of metaphors or analogies, or require modifications in light of new assumptions or empirical data. The phenomenon of the scientific development of concepts is not new. There is a vast body of influential methodological literature on concept formation in the social sciences (Collier, LaPorte, & Seawright, 2012; Gerring, 1999; Goertz, 2006, 2008). The scientific approach to this question has almost always been interested in rigorous methods and this fact appears to be, to some extent, a consequence of the impact of classic research in philosophy of science (e.g. Hempel, 1952) on various fields of science.

The standard theme of concept formation in the (social) sciences is only a part of the broader story about scientific intentional concept development. Current debates on conceptual engineering (CE, henceforth), often understood precisely as intentional or systematic concept development that involves their assessment or evaluation (Cappelen, 2019; Cappelen, Plunkett, & Burgess, 2019) are in interestingly well suited to better understand the extent to which such topics exhaust certain dimensions of concept development.

CE has been one of the central themes or whole research perspectives in contemporary philosophy, which has been attracting attention of many philosophers of language, mind and social change. These are certainly sufficient reasons to take it seriously in light of how social scientists used to approach to the question of conceptual change.

This chapter is intended as a more comprehensive consideration of the relations between the philosophical method of conceptual engineering and intentional concept development in the social sciences[1]. My target is to show that the social sciences exercise specifically understood CE on a daily basis. What they do, it meets basic but essential criteria of CE, although the extent to which this method is philosophical is limited. The limitation does not come from the fact that philosophy is uniquely positioned when it comes to conceptual ameliorations, but from the fact that practices involved in concept development in the social science do not always meet the level of conceptual rigor that would be satisfying in philosophical terms. This opens news ways to conceptual ameliorations in soft sciences, which require cooperation between the social scientists and philosophical conceptual engineers and more complex joint actions that increase the standards of scientific concept development.

To show special relations between CE and intentional concept development in the social sciences, I will focus on the so-called *conceptualizations* and *reconceptualizations* with a special attention to the case of organizational and management research (one of the "softest" areas of the soft sciences). If I am not mistaken, these two approaches or strategies in theorizing and writing papers have gone completely unnoticed in philosophy of the social sciences. Still, I believe that they are interesting enough to shed some light on the practice of concept development and its relation to CE and the philosophy of science in more general. The fact that (re)conceptualizations are not only popular but also rather a standard techniques of writing research papers in organizational research is a reason to treat this field as a representative sample of the social sciences and a good case for discussion of CE in the context of the social-scientific inquiry.

All this affects the way how CE will be handled in this chapter. CE used to be studied from various perspectives: besides the traditional version related to the idea of explication (Carnap, 1950), we usually see philosophy of language (with a special focus on metasemantics) and social philosophy. Here, the perspective will be principally adjusted to philosophy of the social sciences. The issues of linguistic and social change will emerge as filtered through the scientific inquiry.

### Conceptual engineering – the basics

To connect intentional concept development in the social sciences with CE, we need rather minimalistic approach to the latter and some good examples of concepts from the former. Let's begin with the intuitions about CE. It is about "engineering" and about concepts. Take the latter first. Thinking and knowing need concepts. Although there are various approaches to what is or may be engineered (Isaac, 2021;

---

[1] I understand the social sciences broadly as embracing also organizational and behavioral research (with exclusion of behavioral genetics and neurosciences).

Koch, 2021; Koch, Löhr, & Pinder, 2023; Plunkett & Sundell, 2023), I understand CE as about concepts taken as representational devices. Of course, those representational devices may be conceived in various ways and we usually select here between philosophical and psychological views (Isaac, 2023). When it comes to the initial requirements of the social sciences regarding concept development and the way social scientists usually approach concepts or constructs, such decisions do not appear to be essential (cf. Goertz, 2006; Podsakoff, MacKenzie, & Podsakoff, 2016). It does not seem to be crucial, for example, to prefer the psychological view that concepts have interrelated cognitive functions (Solomon, Medin, & Lynch, 1999) over the philosophical view that concepts have epistemic goals (Brigandt, 2010). These views may meet and complement one another (cf. Löhr, 2020), and both philosophy and psychology are certainly helpful when it comes to better understanding what is the conceptual vehicle of scientific message.

Despite some trends to connect CE with psychological views on concepts (Koch, 2021; Machery, 2017), the stake here is not—at least directly—our "cognitive life", but knowledge that already meets certain, scientifically accepted criteria; for this reason, more traditional, philosophical approaches to concepts appear to be still very useful here.

Concepts have structure. Both classic distinctions such as intension/extension from intensional logic (Fitting, 2015) and more subtle issues related to semantic indeterminacy such as "open-texture" (cf. Zayton, 2022)) potentially matter here. To simplify the issue for our minimal requirements regarding CE for the (social) sciences, I will defend the view that the elementary structure of concepts is, primarily, tripartite: concepts have *intension* (sense, meaning), *extension* (scope, range of referents) and *functions* (or play explanatory/epistemic roles) (cf. Brigandt, 2010). Of course, concepts may be disassembled further and other elements (such as e.g. prototype, family resemblance, peripheral features, vagueness) are also of significance for the scientific inquiry. In our context, though, the three-partite structure appears to be crucial.

To illustrate this, let's take two examples: the concept of stratification from sociology [1] and the concept of classical conditioning in psychology [2].

[1] *Stratification* roughly means (*intension*) hierarchical arrangement of individuals or groups in a society. What makes the concept interesting from the scientific point of view is the scope of factors (*extension*) that decide about the arrangement e.g. wealth, income, education, or social status. Interestingly, such scope may also embrace global inequalities or an impact of technology on the arrangement, so it may be broader. The concept of stratification initially explained how individuals are ranked within a society (one of its key *functions*), but it may also help show how

global dynamics contribute to unequal distribution of resources and social chances (see: Grusky & Weisshaar, 2014; Wilson, Wallin, & Reiser, 2003).

Now, let's move to the concept of [2] *classical conditioning* that famously originated from Pavlov's work (Pavlov, 2010). It means (*intension*), roughly, an automatic learning process in which an organism's conditioned behavioral response is a result of associating a neutral stimulus with a meaningful stimulus. The scope of referents (*extension*) of the concept may be broader than the original Pavlovian version comprised, e.g. it may embrace also emotions. The meaning of classical conditioning may also be slightly different when it embraces the issue of timing (so called temporal classical conditioning). Explanatory roles of classical conditioning (*functions*) embrace various issues. One of the roles of this idea is to help explain how automated physiological responses can be conditioned on the basis of linking stimuli with bodily reactions. The functions may also embrace more details such as emotional reactions, their strength or persistence (Allen & Madden, 1985; Black & Prokasy, 1972).

The two concepts briefly introduced above reveal their core three-partite structure with respect to conceptual dynamics of research related to them. The structure can be traced and analyzed in virtually all crucial concepts that play roles in theorizing (modeling, typologies or full-fledged theories). If CE in the social sciences is about concepts, it is about their key structural elements.

Let's now proceed to "engineering". As hinted at the beginning, CE can be understood as intentional or systematic development of concepts based on their assessment. It may be conceived as a certain process (Chalmers, 2020; Isaac, Koch, & Nefdt, 2022) or practice (Löhr & Michel, 2023). "Engineering" in CE is, of course, a metaphor (we are on the terrain of the activity of human mind and, hence, the talk of engineering should not bear more sense than the talk of architecture). It is a metaphor for concept development, i.e. forming new concepts or updating existing ones on the basis of assessment. This view is rather widely accepted in the literature. One of the most important assumptions behind CE is that concepts may be innovative, outdated, suitable, right, wrong, defective, and so on. This implies that conceptual engineers (or developers) assess concepts as having those properties. Thus, concepts can be good or bad with respect to the aforementioned three-partite structure and those properties are gradable (e.g. a concept can have a good scope, but not apt when it comes to its functions, or it can be okay when it comes to its functions, but not all of them; its scope can be too broad, too narrow, and so on). I am not going to discuss this issue further or defend it here on philosophical grounds[2]. For the present purposes, we should distinguish two versions of CE with respect to its goals, which can be more practical or more cognitive or epistemic. Exploiting important contributions of Sally Haslanger (e.g. Haslanger, 2012) and Theodore Sider (Sider, 2011), let me describe them as follows:

> *Haslanger-style CE*: critique of social, legal and political concepts (marriage, gender, race, rape, torture, etc.)—concept development with explicitly practical goals, such as social justice or animal welfare;
> 
> *Sider-style CE*: epistemologically oriented concept development: scientific improvement of the epistemic access to reality and more smoothly gather information about the social and natural world;

---

[2] These issues occupy a larger part of the debate on CE. Many authors agree that engineering meanings is basically possible. This view can be defended in various ways: on the ground of semantic externalism (Koch, 2018), as an empirical view regarding conceptual control (Fischer, 2020), on pragmatic grounds

Although Haslanger's work has been tremendously influential in the social sciences, my understanding of CE gravitates towards the perspective a la Sider. The gravitation can be perceived as a consequence of the view that concepts are representational devices that build scientific world view. Those devices represent the world—be it physical, biological or social. One may accept here various epistemological and metaphysical stances (from hard realism to pragmatism or constructive empiricism), but let me just end on the idea that Sider-style CE in the social sciences is about better and more accurate access to social reality through developing epistemically more optimal concepts. Within this perspective, some psychological, social, legal or organizational concepts may be *scientifically more privileged* when it comes to such access than other concepts of those kinds.

Now, to better understand the ameliorative or optimizing goals of CE when it comes to the access to social reality and the variety of its manifestations, we need a few distinctions. Basically, there are two, equally important versions or dimensions of CE: one is oriented on creation of concepts and the other one—on their revision. More technically:

> *Conceptual engineering\* (CE sensu stricto)*—intentional concept development, i.e. cognitive optimization of the conceptual apparatus in a given domain of knowledge that results in conceptual novelty, new concepts and newly coined terms (conceptual innovations—design or construction *de novo*, e.g. with the use of new metaphors);
> 
> *Conceptual reengineering*—conceptual engineering that is guided by preexisting, but deficient concepts which are semantically corrected or replaced by other

---

(Simion & Kelp, 2019) or, by moving the question to the area of experimental philosophy (Andow, 2020; Nado, 2021).

> concepts in the process of cognitive optimization by means of an explicit characterization (conceptual repairs: revisions and updates);

The two faces of CE are widely recognized in the literature (Brun, 2016; Chalmers, 2020; Prinzing, 2018; Simion & Kelp, 2019). Coining new terms with the aid of metaphors is a standard in the social sciences (metaphors from physics, chemistry and biology have actually been swarming the social sciences (Hodgson, 1995; Langer, 2015; Sousa Fernandes, 2008)). Conceptual updates are no less frequent. Before I illustrate the two versions of CE with examples of concepts and analyze them with respect to the three-partite structure, let's introduce further distinctions. Conceptual *re*engineering may take a few forms, depending on the final result of the process. CE may be:

— *conservative*: after rational assessment (cf. Belleri, 2023), it preserves the current form of a concept (usually: homonymous/same-word CE (Chalmers, 2020));
— *progressive*: after assessment, it updates the concept, either
  a. *modestly* (revision): manipulation in the concept structure to revise or update it,
  b. *radically* (partial or full elimination), by replacement (heteronymous/different-word CE (Chalmers, 2020)) or total rejection of a concept;

Concept development in the social sciences

Now let's put some flesh onto the above presented skeleton. At first glance, it is tempting to claim that various forms of CE are exercised in the social sciences. And CE sensu stricto appears to be as rife as CE understood as revisions or replacements. Let's briefly review this claim; I will begin with the former, i.e. engineering de novo.

Conceptual innovations not only allow us to move research forward but are to significant extent responsible for the fact that science can be also exciting. Good metaphors or analogical thinking in science are important as facilitators of scientific communication. Take for example, [3] the concept of *defense mechanisms* in (psychoanalytic) psychology or [4] *social mirror* in sociology. Both of them are good metaphors. The first one comes from the psychoanalytic theory and it means (intension) mental strategies that are used unconsciously to protect oneself from feelings and thoughts regarding internal conflicts and external stressors (APA, 1994). According to the original (Freudian) version, the defense mechanisms encompass (extension) ten observable strategies: repression, regression, reaction formation, isolation, undoing, projection, introjection, turning against one's own person, reversal into the opposite, and sublimation or displacement. I will not discuss them here. The concept used to play several roles (*functions*), e.g. explains how humans manage inner conflicts, preserve psychological stability or navigate through difficult emotions and situations (Cramer, 1998). Why defense mechanisms are a metaphor, we probably do not need to explain (the mechanisms metaphor is now widely accepted in science and philosophy and it usually captures technical issues—cf. mechanistic explanations).

The concept of *social mirror*, a central concept of social mirroring theory (Whitehead, 2001), generally means (*intension*) that society reflects and shapes human identity and behavior and thus it contributes to the self-conception of individuals. The scope (*extension*) of social mirror embraces, inter alia, cultural values, norms, social roles or shared beliefs. The key goal of the concept (*function*), in the original (Diltheyan) version, is to explain the relation between introspection and "public performance", describing how individuals construct their self-conceptions by aligning their behavior and presentation with social values and norms (Whitehead, 2001).

I tend to think that conceptual novelty and creativity in the social sciences in general share the distinctiveness of the two examples above. For this reason, it is not controversial to actually acknowledge the existence of the narrow sense engineering in the case of conceptual innovations. CE appears to be a rather quick product of scientific creativity and, in fact, it looks easy (Simion & Kelp, 2019). Even if we agree that science is a social practice and it is entangled in complex diachronic processes related to inquiry, conceptual creativity behind metaphors, analogies or innovative conceptual blends appear to be sufficient to accept CE sensu stricto as exercised in the social sciences.

Now let's focus on *reengineering*. Here, the situation appears to be different. Conceptual changes and updates or eliminations (of all kind) are not one-time acts of scientific creativity. Conceptual preservations, eliminations (partial and complete)—operations that prima facie look like good candidates for, respectively, conservative and radically progressive CE—are something that usually takes extended periods of time and many publications (unlike coining new terms or introducing new metaphors). Successful falsifications and other type of critical reconsiderations of concept structure (intention, extension, and key epistemic functions) —potentially leading to partial or complete rejections of certain ideas or theories—require broader acceptance and consensus on the societal level which take us to collective knowledge (Koons, 2022)[3].

---

[3] This appears to be also true in the case of CE sensu stricto, but only when it conceptual innovations are entangled in scientifically more ambitious transformations (see: Thagard, 1992). Conceptual change in the social sciences may appear without those macro-transformations.

To consider a concept preserved, we usually need more than one paper or one book[4]. What is more, these issues are often inherently related to competitive research perspectives, not only between researchers and groups to which they belong (O'Connor & Weatherall, 2018), but also within those groups (cf. Dang, 2019).

To illustrate the issue of preservations, let's briefly consider the concept of diminishing marginal utility (or DMU for short) from the classical economics [5]. Roughly, DMU means the phenomenon that each additional unit of a given resource (e.g. money) leads to an ever-smaller increase in subjective value (Stigler, 1950a, 1950b)). It becomes interesting as a law o DMU: roughly, as given resource wealth increases, the value placed on each unit of the resource decreases. Over time, DMU was attacked from various positions: as a part of utility theory (Myrdal, Swedberg, & Streeten, 2017), as limited (too narrow extension) when it comes to applications (Easterlin, 2005; Veblen, 1909) or from other perspectives (esp. the Marxist positions (cf. Robinson, 2017)). Despite these turbulences, DMU still belongs to the mainstream economics (and it is also considered valid from the vantage point of developmental psychology (cf. Ahl, Cook, & McAuliffe, 2023)).

The case of DMU may be treated as an instance of *conceptual preservation.* This outcome nonetheless is not a one-time act. Crucially, it is not *pre-planned* as a conservation: it involves diachronically extended and socially complex processes of challenging a concept from various perspectives the results of which may be difficult to predict. For these reasons, it is rather problematic to consider such dynamics as conservative reengineering (certainly, one can label it in this way ex post, but this is a completely different matter).

The analogical situation obtains in the case of conceptual *eliminations* (in both versions: full and partial). If they are possible in the social sciences (with respect to entire 3-partite structure)—eliminations also take time and are entangled in similarly complex social and temporal processes. Let's briefly discuss both types of eliminations. *Partial* eliminations, so called *conceptual retirements* (Haueis, 2021), where a given concept changes its status because it fails to serve their central epistemic functions (while its other functions may still be useful for other reasons) shares the same diachronic pattern as conceptual preservations. Haueis studied this issue on the example of the cortical column concept in neuroscience. His analysis of publications reveals that the concept has lost its main role of the neocortical building block, but retained other functions as a guide and "cautionary tale" for research. Conceptual retirements appear to be quite radical in terms of conceptual dynamics, but should not be treated as full eliminations of concepts. Interestingly, they are also visible in the social sciences such as communication studies where scholars discuss the status of such concepts as e.g. [6] media environment (Ewoldsen, 2017).

---

[4] This issue can be analyzed from the perspective of historically oriented philosophy of science (Kuhn, 1962; Lakatos, 1968). Of course, various questions are involved here, including the contexts of justification and discovery (Schickore, 2022; Sturm & Gigerenzer, 2006).

Full eliminations are not quick and they, actually, may take much longer than conceptual retirements. The history of scientific racism (since Enlightenment) and current status of the concept of [7] race—widely rejected as a biological concept—shows that this issue is temporarily and socially too complex (cf. Barkan, 1992; Drescher, 1990) to treat its elimination from the scientific inquiry in terms that are typical for CE. Given the historic context, it would be hard to treat the dynamics of this concept in terms of intentional conceptual development[5].

Considered in broader epistemic dynamics, conceptual eliminations often precede *replacements*. Conceptual replacements also do not seem to be one-time acts when it comes to scientific inquiry. Here, again, we are on the terrain of historically oriented philosophy of science. Such issues have already been studied as changes in entire systems of concepts (Thagard, 1992) or in paradigms (Kuhn, 1962; Lakatos, 1968). From this point of view, the general dynamics of conceptual changes appears to be quite similar in the social sciences and in hard sciences (even if the dynamics of the latter appears to be significantly faster).

The prospects of using the idea of modestly progressive CE to capture some aspects of the dynamics of concepts in the social sciences look slightly more positive. The diachronic scale and complexity of modest conceptual transformations is much less impressive: explicit manipulations in the structure of concepts are frequent in the social sciences. One of the reasons here is the issue mentioned in the introduction—the tempo of social and practical changes. The dynamics of concepts must follow the social and practical dynamics. Small-scale revisions and updates (or "reforms" (Löhr & Michel, 2023)) are thus intended as adjustments or corrections to what turns out to be outdated or obsolete in meaning, scope and concept functions. For this reason, such updates appear to be good examples of modestly progressive CE understood building blocks of incremental progress in given fields of knowledge. The remainder of this section and the next one are meant to show this in more detail.

Let's return to and reconsider our initial examples. We saw that the concept of stratification in sociology [1] may have a broader extension and serve different epistemic functions. Although stratification is a complex phenomenon and it may require various variables in empirical research, considering such factors as globalization in this context changes the concept of stratification and its explanatory roles. One may respond that we are talking here of global stratification—not of social stratification and these are two different concepts (or "constructs"): one refers to society and its layers and the other one—to global (or international) society and its hierarchy. Although such distinctions appear to make sense, global stratification may still be considered an important update or revision of the concept of social stratification. As we learn from the works that discuss this issue in broader theoretical perspective (Wallerstein, 2000), global aspects of stratification are tied to standard social stratification (Grusky, 2011). There is conceptual connection between the two, so one may maintain that that there is a modest conceptual update here.

---

[5] Of course, in current debates on CE the concept of race is subject to philosophical discussion strictly in terms of CE, but this is a different matter and this issue goes beyond my interest in this section.

The concept of classical conditioning in psychology [2] experienced analogical intended updates. We noted that the original version of the concept and newer approaches to it differ in meaning, scope and concept functions. It is not a surprise that there is a shift in meaning of classical conditioning when we move from the classical view of Pavlov who considered the contingent pairings of stimuli (Pavlov, 2010) to e.g. temporal (the timing of stimuli) or emotional versions of conditioning (e.g. conditions of opposing emotional responses (Solomon, 1980)). These shifts show that adding a certain degree of empirical nuance to a given phenomenon leads to gradual shifts in concept structure. Basically, the entire development of research related to classical conditioning can be perceived as a chain of moderate conceptual reengineering procedures that refine and slightly reform the original concept[6].

Let's take stock. I have shortly illustrated the types of CE introduced above with examples from various fields of the social sciences. According to what I attempted to show, temporarily extended, radical forms of conceptual changes do not seem to be good instances of CE, unlike conceptual innovations and conceptual moderate updates. But the latter two appear to be a bread and butter in inquiry. This is probably a bold view. A skeptical reaction to this view might be: why, actually, should we connect such types of conceptual dynamics with conceptual engineering? We know that CE is intentional, so it must be explicitly and deliberately about concepts and their structure. And—it seems—this is not always the case when social scientists forge new concepts or revise the existing ones. Rather, such processes are intuitive, implicit and often driven by existing research practices, so they hardly count as instances of CE. This charge appears to be serious and it is a *methodological challenge* for CE.

Before I address it, I will, firstly, shed more light on the practice of concept development focusing on organizational research.

---

[6] Although some scholars complained here about "little conceptual progress" (Bitterman, 2006), such small-step progress is basically the essence of moderate conceptual updates—not only in this stream of research.

## *(Re)conceptualizations* – the case of organizational research

Probably the simplest way to convince that conceptual innovations and moderate conceptual updates aren't implicit, but are the instances of full-blooded CE would be to show that they are exercised exactly as intentional development of particular deficient or inadequate concepts with special sensitivity to their structure. I am prone to think that this is what we see in organizational research under the tags of so-called *conceptualizations* and *reconceptualizations*. They are rife in this field and often appear in the titles of research papers. Let me give a few random examples what concepts are at stake and of what kind of concept development they are the products. New conceptualizations: risky leadership—a construct built de novo as a result of known connections between the issue of risk perceptions and research on informal leadership (Zhang, Nahrgang, Ashford, & DeRue, 2020), union commitment—a novel concept based on the integration of two different theories about unions of voluntary organizations (Sverke & Kuruvilla, 1995). Reconceptualizations: workplace commitment—a conceptual moderate update that proposes to narrow its extension (Klein, Molloy, & Brinsfield, 2012), absorptive capacity—a conceptual update that proposes to broaden the extension of the concept to reduce ambiguity in empirical studies (Zahra & George, 2002); task complexity—a revision that proposes to change intension of the concept based on the broadened extension (Hærem, Pentland, & Miller, 2015).

The above list of the two types of CE may easily be extended. One may also want to try analyzing other types of CE in this large and heterogenic subfield of the social sciences[7]. If we acknowledge that the field of organizational and management research is also methodologically sensitive to the question of concepts, we should not treat this field as negligible case when it comes to its broader usefulness for the philosophy of science. The work on imported concepts (Oswick, Fleming, & Hanlon, 2011), conceptual blending (Cornelissen & Durand, 2012; Kenttä, 2015), not to mention such classic methods-related topics as conceptual definitions (Podsakoff et al., 2016) or metaphors (Boxenbaum & Rouleau, 2011) convinces that this field is methodologically sensitive to the issue of concept development. Also, the presence of such topics as conceptual clarity (Suddaby, 2010) or parsimony (Shaffer, DeGeest, & Li, 2016) enrich this conviction.

---

[7] This pertains, for example, to conceptual *replacements* and *retirements*. The view that conceptual *replacements* in the social sciences hardly count as radically progressive CE because they take much time could be illustrated with the transition from the concept of *programs* which automatize organizational behavior (March & Simon, 1958) to organizational *routines* (Nelson & Winter, 1982) which are conceptually much richer. "Retirements" are rather similar. Although they can be explicit and intentional—e.g. *grand challenges* (Seelos, Mair, & Traeger, 2023) they appear to share quite similar dynamics as replacements: such attempts at partial eliminations gradually generate reactions from a scholarly community and the final result of this process is difficult to predict.

For these reasons, I believe that many instances of (re)conceptualizations in organizational and management research—that display an understanding of the structure of a concept and some skill in conceptual development—will count as the two kinds of CE described above, conceptual engineering sensu stricto and conceptual moderate reengineering. If this is the case, the view that CE is normally exercised in the social sciences may be reasonably defended. Not only in the case of conceptual innovations—that are "not even distinctively hard" (Simion & Kelp, 2019)—but especially in the case of moderate conceptual updates. Conceptual moderate revisions not only visibly drive incremental progress in research, but also require some technical knowledge about the basic structure of concepts, the way to modify it and good reasons why to do it. These three issues taken together show that (re)conceptualizations are especially interesting for conceptual engineers who like to consider intentional concept development in the context of dynamics of scientific concepts.

The instance of organizational research I exploited seems to be also philosophically interesting for the reason already mentioned a few times. Changes of organizational world are well diagnosed (there is even an acronym for this problem: *VUCA*—volatility, uncertainty, complexity and ambiguity (Bennis & Nanus, 1986)). The fact that the tempo of those changes poses practical problems is one of the key reasons why organizational and management research exists as a subfield of the social sciences: better understanding of quick practical and organizational changes helps dealing with the problems these changes generate. Such understanding requires rather frequent updates in conceptual schemes.

Of course, one may lament that, precisely, for this reason organizational research appears to be chaotic (or even not a science at all). Still, if we want to deal with practical changes in a more systematic or methodic way we probably have no choice but accept the need of revising the concepts we know and want to use[8]. It appears, to a significant extent, that the tempo of conceptual updates is the price that this field pays for being empirically adequate[9].

Other social sciences have their own—often more scientific—reasons for conceptual changes. Still, when it comes to the mechanics of CE, organizational research does not seem to be much different from other social sciences as it piggybacks on the methods and theories of (inter alia) psychology, sociology and economics.

---

[8] This claim refers to many (if not most) systematized types of knowledge of human practice, including philosophical ethical systems. From this stance, one may maintain that some conceptually important aspects of (e.g.) Kant's practical philosophy are *outdated*. Human practice is changeable and philosophical ethics that aims at adequacy is not immune to the challenge of conceptual updates.

[9] I distinguish this issue from the question of conceptual or terminological fashions as well as from the question of purely institutional factors that drive the putative need of conceptual updates. These are different matters that require a separate attention.

Methodological feasibility: (re)conceptualizations as conceptual (re)engineering

Someone may—rightly—respond that the picture of CE in the social sciences painted here is rosy and all this is a way too quick: these instances of conceptual dynamics and developments are not anything like philosophical CE—they only look similar on the surface if we engage in *ex post* reconstruction. True CE is a *normative* enterprise and it should always explicitly reveal authentic reasons for developing concepts in the scientific spirit—to make them not only empirically more adequate, but also rigorous (clearer, more consistent, and so on). This would the lesson from the Carnapian explication. In this context, measuring the practice of concept development in the (social) science in philosophical terms and in accordance to a philosophical method may look unfeasible. Philosophical ameliorative projects appear to be detached from the scientific practice, so they cannot interfere with it, because they are, to a significant extent, alien to the methodological standards and dynamics of concepts in a given field. Or so it seems.

Let's take an example of clarity. It is a well-known desideratum for concept formation and development in the social sciences. For a philosophically trained conceptual engineer it may be very tempting to increase the standards of clarity in a given field of knowledge: the analytic philosophers know all too well that theories in the social sciences often suffer from the lack of clarity and precision[10]. It may be, therefore, tempting to philosophically *repair* (or even *eliminate*) certain concepts. But in normal circumstances—where philosophers and scientists do not do their thing together—such repairs (or eliminations) may be perceived as a kind of epistemic trespassing (Ballantyne, 2019) or conceptual domination (Shields, 2021). Why actually a sociologist—especially the one who does not like nuance (see: Healy, 2017)—should care what armchair philosophers try to tell them about clarity?

In debates around CE, such a criticism is very closely related to *the implementation problem*. Interestingly, this problem usually is not discussed in the context of scientific inquiry (see, e.g.: Jorem, 2021; Queloz & Bieber, 2021). To deal with it in our context—and to argue for the special bond between CE and (re)conceptualizations—I will distinguish between two approaches to CE. The first one treats CE as a certain practice (following Althusser one may call it a "theoretical practice" (Althusser, 1969)) and a form of actual scientific inquiry. The second one treats CE in terms of a more systematic research program. Only the first approach sanctions the identification of intentional concept development in the social sciences with CE.

---

[10] The problem has become more visible after the publication of *Fashionable Nonsense* (Sokal & Bricmont, 2003), but it still has not been solved; there are subfields of the social science and journals that are harbors of what Cappelen and Dever call "deep bullshit" (Cappelen & Dever, 2019).

Again, the classic themes in philosophy of science related to scientific *rigor*, allow one to show why (re)conceptualizations should be treated as conceptual (re)engineering understood as a theoretical practice. Rigor can take various forms. For our purposes, it will be sufficient to associate it with conceptual clarity, consistency or precision. I mentioned that these issues are well known as desiderata for concept formation in the social sciences. Although they are widely shared, we observe also other tendencies. Those tendencies may be summarized in Quine's dictum: *we may prudently let vagueness persist* (Quine, 1960, p. 116). If we consider this issue more seriously, it turns out the use of vague and ambiguous terms may have surprising benefits for research—despite all the positivist ideals of rigorous science.

The issue is not new. It was studied in the methods of behavioral sciences more than a half of the century ago. In his well-known handbook Abraham Kaplan noted: "Tolerance of ambiguity is as important for creativity in science as it is in anywhere else" (Kaplan, 1964, p. 71). Conceptual inconsistencies and unclarities may be beneficial for increasing empirical adequacy in such fields as anthropology (Djordjevic, 2020) or they may help integrate certain forms of knowledge, e.g. in biology (Neto, 2020). The list of those benefits may be extended if we consider research practices in the social sciences. All this suggests that the actual conceptual dynamics of various forms of scientific knowledge differs when it comes to its standards of rigor from what we know as desiderata for concept formation or analytical values that should guide research. Or they integrate those standards in their own ways. Various subfields of scientific knowledge in the social sciences have their own, local and often only implicit criteria for acceptable degree of rigor as much as they have their own understandings of unacceptable jargon or overly abstract terminology (scholars who work in one field and try to submit a research paper to a journal from a different field will understand this problem). Those criteria are frequently an upshot of local academic practices in those subfields. They are dependent on their own histories and epistemic cultures which is the reason why they have, to an extent, their own logics and conceptual dynamics. Sociology of science understood through the lens of locally parceled practices and cultures already investigated this topic (Knorr Cetina, Schatzki, & von Savigny, 2001; Knorr-Cetina, 1999).

Of course, the above sociological picture of disunity of inquiry is not the whole story about CE. At least—it should not be. The fact that intentional concept development must meet basic criteria of inquiry when it comes to conceptual novelty and updates, provides an anchor for conceptual engineering understood as a *research* *program*. Although they are locally parceled out and highly heterogenous, social sciences nonetheless gravitate towards conceptual rigor (no matter how chaotic their instances sometimes are). The theoretical practice of (re)conceptualizations may always look better and this is why the analytically refined, ameliorative projects adjusted to the standards of inquiry in the social sciences make sense epistemologically. Analytical philosophy and social sciences can be epistemic neighbors (Watson, 2022). This takes us again to the question of desiderata for concepts but, this time, adjusted explicitly to the needs of CE behind (re)conceptualizations. Here I would like to formulate a few basic recommendations for concept developers in the social science and then briefly describe ways how to implement them.

First, it seems that the knowledge of the basic structure of concepts should be much more common among scholars who seek novelty or want to update concepts. This is not very difficult and will help making conceptual development more standardized.

Second, any attempt to create a new concept or reconsider an existing one should be careful. Elsewhere, I proposed to refresh and broaden a set of standard guidelines for concept development (Makowski, 2021). Except for concept formation based on *clarity*, *parsimony* or *validity*, the very process of updating concepts should be more rigorous. Here are a few of the values that may help:

(1) *soundness*: if concepts are imported from other fields or theories, their meaning and functions should not be distorted in the process of domestication.
(2) *reflexivity*: reasons for conceptual revision should always be explicitly weighed.
(3) *transparency*: the intention to edit a concept should always be explicitly communicated and understandable for the audience.

Accepting these values in the practice of (re)conceptualizations would amount to embarking on CE as a research program. How to do it? I see three paths that potentially cross.

The first path would be to show the bridge between CE as a practice and CE as a research program by pointing to the idea of *best practice recommendations*—a well-tested solution in the research methods literature (cf. Bretschneider, Marc-Aurele, & Wu, 2004; Osborne, 2008). Best practice recommendations for CE would be guidelines built on the basis of the above epistemic values which reflect the standards of inquiry (plus expert consensus about those standards) supported by proper evidence of actual exemplary (re)conceptualizations. They would provide sufficient epistemic ground for the continuity of inquiry.

The second path is institutional. This path has already been initially explored under the tag of *the politics of implementation* (Queloz & Bieber, 2021). In the context of inquiry, this idea does not directly involve the question of big politics (e.g. democracy), but it amounts to building stronger institutional ties between philosophy (especially: of science) and the social sciences. Here it is entirely inquiry-related (a consequence of the view that CE is understood here within the bounds of the scientific inquiry—introduced at the beginning), so it is more a question of policy-making. Hard scientists already entered this path. Laplane et al. (2019) already described the most important actions and we may easily adjust their recommendations for the social sciences. These recommendations amount to the increased presence of analytic philosophy in institutionalized research in all subfields of the social science—from teaching to publications.

There is also a third path that emerged just recently: the use of AI-agents in conceptual work. It is promising both as a potential booster of de novo conceptual engineering and as an aid in revising existing concepts thanks to new conceptual associations (without risking too much jargon or nonsense). Generating original meaningful conceptual content that meets the standard desiderata for concepts is no longer the domain of humans—even when it comes to philosophical content (see: Schwitzgebel, Schwitzgebel, & Strasser, 2023)—and progressive conceptual engineers should not neglect this fact.

## Conclusion

I attempted to show what it may appear to be impossible: how often suboptimal conceptual dynamics in the social sciences can be harnessed as sufficiently rigorous conceptual engineering. The attempt proposed to narrow conceptual engineering down to scientific inquiry (Sider-style CE) and distinguished CE as a theoretical practice and CE as a research program. I focused on the so-called *(re)conceptualizations* in the social sciences. As I have shown (I hope), from the point of view of the scientific practice we have reasons to accept the idea that social scientists exercise CE in two versions: CE de novo is employed usually as new *conceptualizations* and moderately progressive CE—as *reconceptualizations*. Importantly, the second type of CE—neglected in philosophy of the social sciences—appears to be highly important for conceptual dynamics and incremental progress of any field of (social) sciences. Still, both types appear to be equally significant also for CE understood as the ameliorative research program and for its prospects in the social sciences. In this respect, I have pointed to three possible paths that remain open and bridge the gap between actual theoretical practices in the social sciences and normative, programmatic approaches to CE: best practice recommendations, institutional actions and uses of AI-agents. Each of them (separately or together) may improve the scientific condition of (re)conceptualizations.